%% file: saner_main.tex
\documentclass[10pt,conference]{IEEEtran}
\IEEEoverridecommandlockouts

\usepackage{cite}
\usepackage{amsmath,amssymb,amsfonts}
\usepackage{algorithmic}
\usepackage{graphicx}
\usepackage{textcomp}
\usepackage{xcolor}
\def\BibTeX{{\rm B\kern-.05em{\sc i\kern-.025em b}\kern-.08em
    T\kern-.1667em\lower.7ex\hbox{E}\kern-.125emX}}

\usepackage[utf8]{inputenc}
\usepackage[english]{babel}
\usepackage[bookmarks=false]{hyperref}
\usepackage{graphicx}
\usepackage{tabularx}
\usepackage{booktabs}

\usepackage{latexsym}
\usepackage{mathrsfs}

\usepackage{enumerate}
\usepackage{xcolor}
\usepackage{colortbl}
\usepackage{longtable}
\usepackage{tabu}
\usepackage{xspace}
\usepackage{balance}
\usepackage{multirow}
\usepackage{epstopdf}
\usepackage{paralist}
\usepackage{ifthen}
\usepackage{amssymb}
\usepackage{amsmath}
\usepackage{nccmath}
\usepackage{enumitem}
\usepackage{fancyvrb}
\usepackage{adjustbox}

\usepackage{filecontents}
\usepackage{listings}

\usepackage{caption}

\newcommand{\ie}{\emph{i.e.},\xspace}
\newcommand{\eg}{\emph{e.g.},\xspace}

\newcommand{\etal}{\emph{et al.}\xspace}

\newcommand\datamaterial{https://doi.org/10.5281/zenodo.3588501}

\newcommand\analyzedFiles{365\xspace}
\newcommand\analyzedReviews{262\xspace}

\definecolor{gray50}{gray}{.5}
\definecolor{gray40}{gray}{.6}
\definecolor{gray30}{gray}{.7}
\definecolor{gray20}{gray}{.8}
\definecolor{gray10}{gray}{.9}
\definecolor{gray05}{gray}{.95}

\newlength\Linewidth
\def\findlength{\setlength\Linewidth\linewidth
  \addtolength\Linewidth{-4\fboxrule}
  \addtolength\Linewidth{-3\fboxsep}
}

\definecolor{maroon}{rgb}{0.5,0,0}
\definecolor{darkgreen}{rgb}{0,0.5,0}
\lstdefinelanguage{XML}
{ 
  basicstyle=\footnotesize\ttfamily,
  morestring=[s]{"}{"},
  morecomment=[s]{?}{?},
  morecomment=[s]{!--}{--},
  commentstyle=\color{darkgreen},
  moredelim=[s][\color{black}]{>}{<},
  moredelim=[s][\color{red}]{\ }{=},
  stringstyle=\color{blue},
  identifierstyle=\color{maroon}
}
\graphicspath{ {figs} }

\newenvironment{rqbox}{\par\begingroup
  \setlength{\fboxsep}{5pt}\findlength
  \setbox0=\vbox\bgroup\noindent
  \hsize=0.95\linewidth
  \begin{minipage}{0.95\linewidth}\normalsize}
  {\end{minipage}\egroup
  \textcolor{gray20}{\fboxsep1.5pt\fbox
    {\fboxsep5pt\colorbox{gray05}{\normalcolor\box0}}}
  \endgroup\par\noindent
  \normalcolor\ignorespacesafterend}

\newenvironment{examplebox}{\par\begingroup
  \setlength{\fboxsep}{5pt}\findlength
  \setbox0=\vbox\bgroup\noindent
  \hsize=0.95\linewidth
  \begin{minipage}{0.95\linewidth}\normalsize}
  {\end{minipage}\egroup
  \textcolor{gray20}{\fboxsep1.5pt\fbox
    {\fboxsep5pt\colorbox{gray05}{\normalcolor\box0}}}
  \endgroup\par\noindent
  \normalcolor\ignorespacesafterend}

\newenvironment{resultbox}{\par\begingroup
  \setlength{\fboxsep}{5pt}\findlength
  \setbox0=\vbox\bgroup\noindent
  \hsize=0.95\linewidth
  \begin{minipage}{0.95\linewidth}\normalsize}
  {\end{minipage}\egroup
  \textcolor{gray20}{\fboxsep1.5pt\fbox
    {\fboxsep5pt\colorbox{white}{\normalcolor\box0}}}
  \endgroup\par\noindent
  \normalcolor\ignorespacesafterend}

\begin{document}

\title{On The Effect Of Code Review On Code Smells}

\author{
	\IEEEauthorblockN{Luca Pascarella}
  \IEEEauthorblockA{\textit{Delft University of Tecnhology}\\
    Delft, The Netherlands \\
		l.pascarella@tudelft.nl}
\and
	\IEEEauthorblockN{Davide Spadini}
  \IEEEauthorblockA{\textit{Software Improvement Group} \\
  Amsterdam, The Netherlands \\
	dspadini@sig.eu}
\and
	\IEEEauthorblockN{Fabio Palomba}
  \IEEEauthorblockA{\textit{University of Zurich} \\
  Zurich, Switzerland \\
  palomba@ifi.uzh.ch}
\and
	\IEEEauthorblockN{Alberto Bacchelli}
  \IEEEauthorblockA{\textit{University of Zurich} \\
  Zurich, Switzerland \\
	bacchelli@ifi.uzh.ch}
}

\maketitle

\input{front/abstract}
\input{front/keywords}

\input{sections/introduction} 
\input{sections/related}

\input{sections/study_design}
\input{sections/results} 
\input{sections/discussions} 
\input{sections/conclusions}

\section{Acknowledgment}
\noindent
A. Bacchelli and F. Palomba gratefully acknowledge the support of the Swiss National Science 
Foundation through the SNF Projects No. PP00P2\_170529 and PZ00P2\_186090. 
This project has received funding from the European Union’s H2020 programme 
under the Marie Sklodowska-Curie grant agreement No. 642954.

\balance
\bibliographystyle{ieeetr}       
\bibliography{references}   

\end{document}

%% file: front/abstract.tex

\begin{abstract}
	Code smells are symptoms of poor design quality. Since code review is a process that also aims at improving code quality, we investigate whether and how code review influences the severity of code smells. In this study, we analyze more than 21,000 code reviews belonging to seven Java open-source projects; we find that active and participated code reviews have a significant influence on the likelihood of reducing the severity of code smells. This result seems to confirm the expectations around code review's influence on code quality. However, by manually investigating \analyzedFiles cases in which the severity of a code smell in a file was reduced with a review, we found that---in 95\% of the cases---the reduction was a \emph{side effect} of changes that reviewers requested on matters unrelated to code smells. Data and materials [\href{\datamaterial}{\datamaterial}].
\end{abstract}

%% file: front/keywords.tex

\begin{IEEEkeywords}
	Code Smells; Code Review; Empirical Studies.
\end{IEEEkeywords}


%% file: sections/introduction.tex

	\section{Introduction}
	\label{sec:introduction}

		Code smells are sub-optimal design decisions~\cite{fowler:1999}. 
		Our research community has found empirical evidence of the negative impact of code smells on software maintainability~\cite{palomba2017diffuseness,olbrich2009evolution,khomh2012exploratory,Spadini:icsme18}, program comprehensibility~\cite{abbes2011empirical}, and development effort~\cite{sjoberg2013quantifying}, as well as on the problems that developers encounter when dealing with these smells~\cite{tufano2017when,tufano2016empirical,peters2012evaluating,bavota2015experimental,palomba2017scent,palomba2018beyond}.

		At the same time, our community has found that code smells are generally not a primary concerns for practitioners~\cite{YamashitaM12,yamashita2013developers,palomba2014they,taibi2017developers,silva2016we,palomba2017exploratory,kim2014empirical}. Morales \etal reported an exception to this behavior~\cite{morales2015code}: They found initial evidence suggesting that developers may be concerned with code smells, but only \emph{during code review} (a development phase not investigated by previous studies). Analyzing Qt, VTK, and ITK projects, Morales \etal found that the classes that receive better code review treatment are less prone to the occurrence of code smells. Indeed, this finding is in line with previous research that has shown that developers during code reviews are not only worried about finding defects but also about general code improvements and design quality~\cite{bacchelli:2013}. Interestingly, the existence of a relation between code review and code smells could indicate an important context in which researchers on code smells can focus their attention (\eg to have the most impact with their tools on detecting and refactoring smells).

		The method used by Morales \etal~\cite{morales2015code}, however, did not allow the authors to establish a causal relationship between code review and code smells. Despite the authors carefully considered confounding factors with their models (\eg size, complexity, ownership), it is impossible to exclude that other factors may be the actual cause of the measured relationship.
		
		In this study, we continue on this line of work and propose a study whose goal is to investigate \emph{the effect of code review on code smells}. We use a novel research method to reach stronger conclusions about the causality of the relationship between code review and code smells: \begin{inparaenum}[(1)]
			\item we consider a fine-granularity (\ie file-level, as opposed to component level)
			\item compare the code smells in each file \emph{before} it enters a code review vs. \emph{after} it exits from it (as opposed to looking at aggregate values, \eg on code review coverage),
			and 
			\item manually inspect cases in which the severity of code smells decreases in a review.
		\end{inparaenum} 
		
		Overall, we \begin{inparaenum}[(1)]
			\item collect data of 21,879 code reviews from seven Java software projects,
			\item measure any decrease in severity concerning six selected code smells in the files under review, 
			\item use statistical analysis to investigate the relationship between any severity decrement to the activity in the corresponding code review,
			\item assess whether specific code smells are more often decreasing in severity within a code review,
			and
			\item manually analyze \analyzedFiles cases in which the severity of a code smell in a file decreased in a code review.
		\end{inparaenum}
		
		Our results (1) indicate that active and participated code reviews have a higher chance of reducing code smells, (2) fail to confirm a relationship between what developers reported as more critical code smells in previous studies~\cite{palomba2014they,taibi2017developers} and code review activity, and (3) reveal that most of the changes in code smells are not due to a direct comment on design issues, rather a side effect of code changes requested by the reviewers on matters unrelated to code smells.

		Overall, the main contributions of this work include:

	\begin{enumerate}
	
		\item Empirical evidence, based on a large-scale study, that active and participated code reviews have the potential of reducing the impact of code smells on reviewed files;
				
		\item Empirical evidence on the types of code smells whose severity is most likely to be reduced because of code review, which includes the discussion of contrasting results with respect to previous findings achieved in the field of code smell perception and refactoring;
		
		\item A classification clarifying the causes that led to code smell severity decrements, whose results highlight that only rarely reviewers actively stimulate the discussion on how to make the design of a class more functional;
		
		\item An online appendix~\cite{appendix}, including all data and scripts used to conduct our study.
	
	\end{enumerate}

%% file: sections/related.tex
\section{Background and related work}
	In this section, we report the background information on the topic of interest and connected related work.

\begin{figure*}
    \centering
    \includegraphics[width=\textwidth]{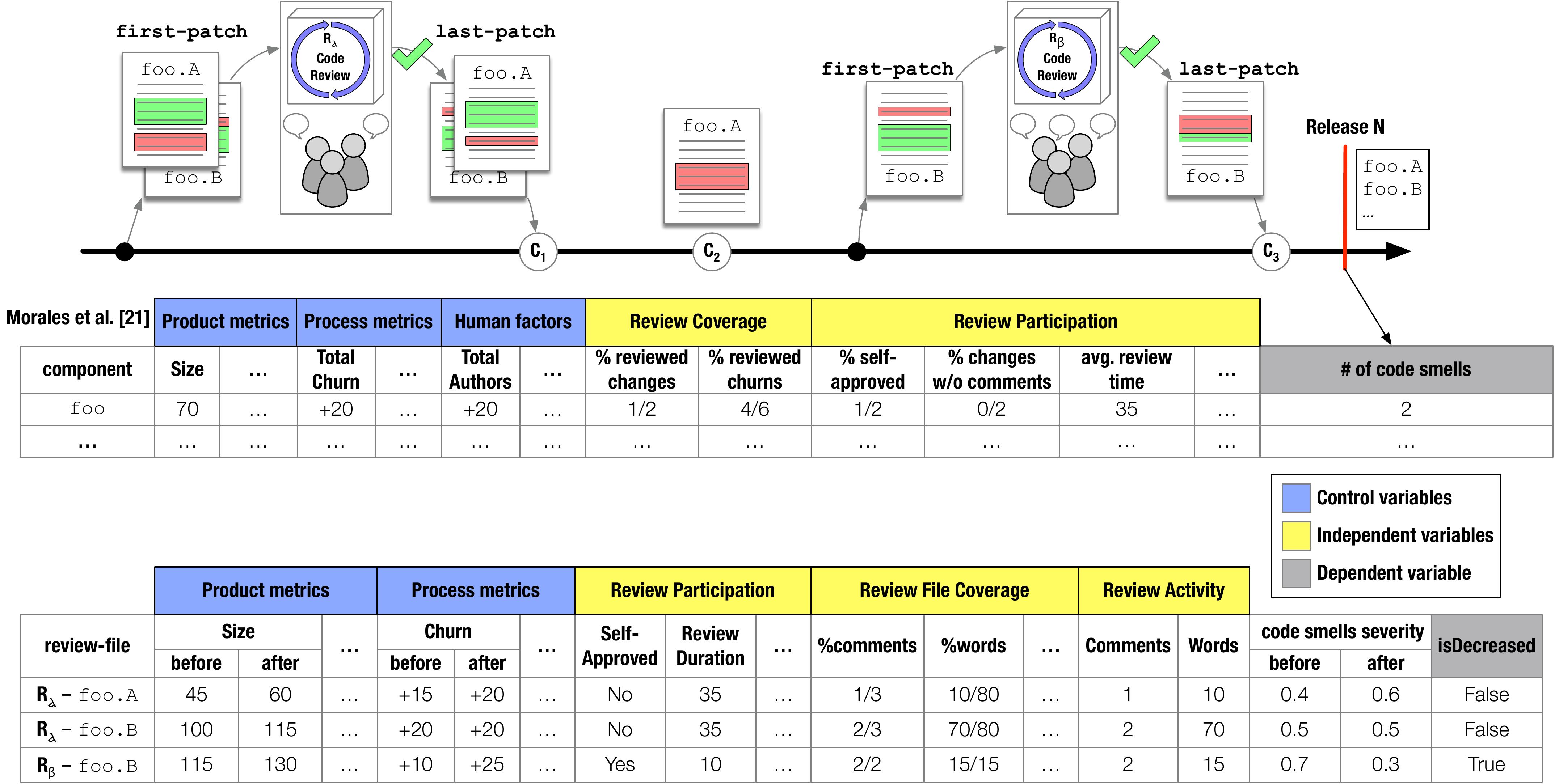}
    \caption{A comparative explanation of how Morales \etal differ from our study.}
    \label{fig:comp}
\end{figure*}

\subsection{Related Work}
\label{sec:related}
In the following we contextualize our work with respect to literature on code review and code smells.\smallskip
	
		\textbf{Code Review.} The research community has extensively investigated how code review activities, particularly participation and coverage of the process, influence source code quality. Abelein~\etal~\cite{Abelein2015} demonstrated the impact of user participation and involvement on system success. 
		Thongtanunam~\etal~\cite{Thongtanunam2015,thongtanunam2016revisiting} found that reviewing participation and expertise can decrease defect-proneness, while Rigby~\etal~\cite{Rigby2014} reported that review participation is the most influential factor influencing code review efficiency. Moreover, studies indicated that patches should be reviewed by at least two developers to maximize fault identification~\cite{porter1998comparing,Sauer2000,kemerer2009impact,Rigby2013,kononenko2015investigating,di2016security,Thongtanunam2016,kononenko2016code}.
		
		Bavota and Russo found that patches with a low number of reviewers tend to have a higher chance of inducing new defects~\cite{Bavota2015}. Furthermore, McIntosh~\etal~\cite{mcintosh2014impact,Mcintosh2016} measured the extent to which the proportion of reviewed patches and the amount of participation in a module impact software quality, finding that a low percentage has a negative impact on code quality. Nevertheless, these findings were later partially contradicted by Sadowski~\etal~\cite{sadowski2018modern} in the case of Google. Ram \etal \cite{ram2018makes} further suggested that reviewers should be involved in the inspection of small patches to be more efficient, while Pascarella \etal \cite{pascarella2018information} identified a set of information needs that would help developers while reviewing code. 
		
		M{\"a}ntyl{\"a} and Lassenius~\cite{mantyla2009types} conducted a study aimed at investigating what types of defects are discovered in code review, finding that, in addition to functional defects, code reviews find many evolvability defects, (\ie issues related to the design and non-functional quality of the code). Later, Beller \etal~\cite{beller2014modern}, as well as Bacchelli and Bird~\cite{bacchelli:2013}, confirmed similar results in different contexts.\smallskip

		\textbf{Practitioners And Code Smells.}
		Researchers have investigated whether and how developers perceive code smells as an issue. 		Arcoverde \etal~\cite{Arcoverde:WRT2011} studied how practitioners react to the presence of code smells in the source code. Their primary finding indicates that code smells tend to live in the source code for a long time because developers avoid refactoring to prevent unexpected consequences. In an industrial environment, Yamashita and Moonen~\cite{yamashita2013developers} reported the results of an empirical study aimed at evaluating the code smell severity perceived by developers of commercial software. The authors found that 32\% of the interviewed developers overlook code smells; therefore, the removal is not in their priority. Confirming this trend, Sjøberg \etal~\cite{sjoberg2013quantifying} found that code smells pose only secondary design problems, thus justifying how developers tend to ignore their presence. Opposite findings were reported by Palomba \etal~\cite{palomba2017scent}, who found that the previous observation was not entirely correct when considering a subclass of textual code smells only. Developers tend to diagnose better design problems characterized by textually-detectable code smells rather than those of structurally-detectable ones. These results are aligned with a previous study where Palomba \etal~\cite{palomba2014they} surveyed a pool of practitioners belonging to three different open source systems and reported that developers consider code smells associated with complex and lengthy source code harmful for maintainability. At the same time, whether the surveyed developers perceived the other code smells as a problem strongly depended on the severity of the problem.

\subsection{Comparison with Morales \etal~\cite{morales2015code}}
		
Morales \etal~\cite{morales2015code} are the first who investigated the relation between code review and code smells. They found that \emph{software components with limited review coverage and participation are more prone to the occurrence of code smells} compared to components whose review process is more active.

Our study has been greatly inspired by Morales \etal~\cite{morales2015code}. We aim to determine a more reliable causal relationship between code review and code smells, by using a novel method. In the following we detail the differences between our study and the one by Morales \etal~\cite{morales2015code}, by considering the scenario in a software system's evolution depicted in Figure~\ref{fig:comp}.\smallskip

\textbf{A scenario of a software system's evolution.} In Figure~\ref{fig:comp}, the horizontal line represents an interval in the evolution of a software system: This interval depicts three commits ($C_1, C_2, C_3$), two code reviews ($R_\alpha$ and $R_\beta$) and a software system's release ($N$, represented by the vertical red line).
First, a developer modifies two files (\texttt{A} and \texttt{B}) that belong to the same component (\texttt{foo}) based on the current state of the main branch and submits the patch (\texttt{first-patch}) for review ($R_\alpha$). The review includes three reviewers who provide two comments over the course of an unspecified number of iterations. During the review, the patch is changed until the reviewers are eventually satisfied with the final state of the patch (\texttt{last-patch}) and allow the author to merge it into the main branch of the system (commit $C_1$).
In commit $C_2$, a developer commits changes to \texttt{foo.A} to the main branch directly, without a review.
Finally, a developer takes the current state of the main branch (\ie as it is after $C_2$) and modifies \texttt{foo.B}; then, the changed file (\texttt{first-patch}) is sent to review ($R_\beta$); and the final version of the file after the review loop (\texttt{last-patch}) is committed ($C_3$) to the main branch.\smallskip

\textbf{Approach by Morales~\etal~\cite{morales2015code}.} Considering the aforementioned scenario, the approach by Morales~\etal considers one of the available releases ($N$) and takes all the released source code files (in our case, we focus on \texttt{foo.A} and \texttt{foo.B}). Subsequently, the authors compute the code smells for these released files, aggregate the information at component (\ie package) level, and consider the result as their \emph{dependent variable}. Furthermore, the authors compute control variables in the form of product (\eg lines of code, at the moment of the release) and process (\eg ownership and churn, over the history) metrics~\cite{rahman2013and}. Finally, Morales~\etal compute review metrics regarding coverage (\eg the proportion of changes that have been reviewed) and participation (\eg the proportion of changes without discussion) as independent variables.
Following this approach, they consider the whole set of changes and commits that affect the evolution of the project between two releases---including the ones where developers did not perform a code review---to compute code review metrics. Although this effect is balanced out by two independent variables (\ie `proportion of reviewed changes' and `proportion of reviewed churn'), this methodology allowed the authors to only obtain a coarse-grained overview of the relationship between code review practices and software design quality. For instance, in the scenario of Figure~\ref{fig:comp}, changes that do not pass through code review (\eg \texttt{foo.A} in $C_2$) may increase/decrease the number of code smells independently from code review activities.\smallskip

\textbf{Our proposed alternative.} In our study, we propose a novel method to better capture the causal link between code review and code smells. First, we consider classes (\ie files) instead of components (\ie directories), and we only consider code changes inspected in a code review. Then, we compute code smell severity \emph{before} vs. \emph{after} each code review (\ie we compute code metrics considering only changes in the \texttt{first-patch} and the \texttt{last-patch}). As depicted in Figure \ref{fig:comp}, our dependent variable is named \texttt{isDecreased}: it indicates whether the severity of a smell has decreased after the code review and is computed computing the difference between the severity of the smell before and after a code review. Changes like the one leading to $C_2$ are not considered by our approach, because they have no direct relationship with any code review activity.

%% file: sections/study_design.tex

\newcommand{\rqone}{How does code review influence the severity of code smells overall?}
\newcommand{\rqtwo}{How does code review influence the severity of the different types of code smells?}
\newcommand{\rqthree}{Why does the severity of code smell(s) decrease during code review?}

\section{Methodology}
\label{sec:methodology}
    Our \emph{goal} is to investigate the causal link between code review and code smells, with the \emph{purpose} of understanding whether the inspection activities performed by developers reduce the severity of code smells. The \emph{perspective} is of both researchers and practitioners, who are interested in gaining a deeper understanding of the effects of code review.
        
        \subsection{Research Questions}
            First, we investigate to what extent the severity of code smells is reduced by the code review process:
            
            \begin{center}
                \begin{rqbox}
                    \textbf{RQ$_1$.} \rqone
                \end{rqbox}
            \end{center}        
    
            Then, we analyze whether some types of code smells exhibit a higher likelihood to be reduced with a code review. Since past studies found that developers perceive certain code smells as more relevant~\cite{palomba2014they,taibi2017developers}, we investigate whether these smells are removed with a higher likelihood.
    
            \begin{center}
                \begin{rqbox}
                    \smallskip
                    \textbf{RQ$_2$.} \rqtwo
                \end{rqbox}
            \end{center}        
            
            Finally, we aim to qualitatively understand why the code smell severity changes with a code review (\eg do the reviewers explicitly request the author to manage code smells?):
            
            \begin{center}
                \begin{rqbox}
                    \textbf{RQ$_3$.} \rqthree
                \end{rqbox}
            \end{center}        
            
\subsection{Context of the Study}

To collect data for our study, we leverage \textsc{CROP} (Code Review Open Platform), \ie a publicly available platform and dataset developed by Paixao~\etal~\cite{Paixao2018}. This dataset is ideal for our research goals, because it (1) stores all review details, (2) contains a large amount of data on projects with different size and scope and developed by different communities, and (3) is fully compliant with the European Union General Data Protection Regulation~\cite{voigt2017eu} (randomly generated data replace real names and email addresses of developers).\smallskip

    \textbf{Subject systems.} Among the 11 software systems available in CROP, we focus on the seven ones written in Java.\footnote{Our research community has more widespread and better validated code smells detection strategies for Java than for other languages (\eg those defined by Lanza and Marinescu~\cite{Lanza:2006}), thus we prefer Java to avoid uncertainties caused by instruments that have not been previously evaluated.}
    The first two columns of Table \ref{tab:projects} report information on the subject software systems, including their names and the time span for which we have data.\smallskip
    
    \textbf{Subject code reviews.} The dataset available in CROP includes the whole suite of patch-sets with the related discussions for each code review in the considered systems. It also gives easy access to the first-patch and last-patch sets (Figure~\ref{fig:comp}) that represent a two-fold snapshot of the source code before and after the code review. Having this information at disposal is a key advantage for our study, because we can measure the severity of code smells before vs. after each code review, thus we can address our research questions. The last two columns of Table~\ref{tab:projects} report the considered code review data.\smallskip
    
    \textbf{Subject code smells.} We consider six different types of code smells, which are reported in Table \ref{tab:smells}. We selected these code smells because: (1) they have different characteristics (\eg complexity vs. design compliance), thus allowing us to analyze a wider range of design issues that frequently occur in real software systems~\cite{palomba2017diffuseness}, and (2) they have already been considered by previous work concerning their relevance for developers~\cite{palomba2014they,taibi2017developers,yamashita2013developers}, therefore they can be used to assess the role of developers’ perception within code review activities.

\input{tabs/tab-projects}
    \input{tabs/tab-smells}

\subsection{Detection Of Code Smells And Their Severity}
A key measure in our study is the variation in the severity of the code smells in classes before they enter the review process and after they exit from it. In other words, we are interested in deriving relative measures such as the coefficient of variation for the severity between two events rather than obtaining an absolute value~\cite{saaty1993relative}. To that end, we first need to detect the instances of the six considered code smells. We use a metric-based code smell identification strategy \cite{pecorelli2019comparing} and, particularly, detection rules inspired to those defined by Lanza and Marinescu \cite{Lanza:2006}: we compute object-oriented metrics \cite{CKMetrics} characterizing the classes that were subject to code review under different angles such as size, complexity, and coupling, then define thresholds that discriminate code artifacts affected (or not) by specific types of code smells. 

After the code smell identification phase, we compute the code smell severity variation ($\Delta severity$) by comparing the extent to which the smelly instance exceeds the given threshold before and after a code review, as in the following formula:

\begin{equation}~
    \begin{array}{l}
        severity(smell, class) = \frac{metricsValue_{smell}}{threshold(smell)}\\ 
        \\
        \Delta severity = severity_{after\_review} - severity_{before\_review}
    \end{array}
\end{equation}

Negative values of $\Delta severity$ indicate that the severity of a code smell decreased as a result of the code review process, while positive values mean that it increased. 

\smallskip
\textbf{Thresholds.} The detection mechanism applied can output some false positives, \ie classes erroneously classified as code smells: should the number of false positives be high, the results of our study would be compromised.
To mitigate this issue, we initially select the detection rules used in previous literature~\cite{fernandes2016review}; this also includes the thresholds adopted to discriminate between smelly and non-smelly artifacts~\cite{garcia2016improved,ferenc2014source,lopez2005relevance,fontana2015towards}. After selecting the detection rules, we manually analyzed some code smell instances given as output of the detection. Therefore, we realized that the thresholds were too relaxed (resulting in too many false positives). To solve this problem, we applied more strict rules, \ie we increased the thresholds for the cyclomatic complexity, or LOC, etc. This tuning allowed us to capture slight variations of $\Delta severity$ reducing false positives. Our appendix reports the entire list of our thresholds~\cite{appendix}.

        \subsection{\textbf{RQ$_1$}. Methodology}    
        \input{tabs/tab-independentVariables}
        To address \textbf{RQ$_1$}, we investigate whether the decrement in the severity of the code smells in the classes submitted for code review vs. the same classes after the code review is influenced by metrics related to engagement and activity in the code review, while controlling for other characteristics. The following subsections report the methodology adopted to define and compute dependent, independent, and control variables as well as the statistical modeling we used. The granularity of our model is class-level: We compute each variable for every class that enters a review and exits from it (\eg as depicted in the bottom table in Figure~\ref{fig:comp}).
            
            \smallskip
            \begin{description}[leftmargin=0.0cm]
                \item[Dependent Variable.] Our dependent variable is a binary variable \textsl{`isDecreased'} that takes the value `True' if the severity of at least one code smell (existing in the file before the review) is decreased after the review and `False' otherwise.
                
                \smallskip
                \item[Control Variables.] Some factors may affect the outcome if not adequately controlled. Since we were inspired by the study of Morales \etal~\cite{morales2015code}, whenever feasible, we have used control variables as similar as possible to the ones used in the study by Morales \etal~\cite{morales2015code}. The first five rows of Table \ref{tab:independentVariables} describe the families of variables we use as \textit{control variables} in our statistical model, (\ie \textsl{`Product'} and \textsl{`Process'} metrics~\cite{rahman2013and}).

                \smallskip
                \item[Independent Variables.] Not all code reviews are done with the same care. Indeed, past research has defined proxy metrics to define engagement and participation in code review and has shown that higher values in these metrics related to a better outcome for a code review~\cite{mcintosh2014impact}. In this study, we proceed similarly: we selected as independent variables a set of metrics to capture the engagement as well as activity during code review and we expect that code review with higher corresponding values will lead to a stronger effect on code smell reduction. The second part of the rows in Table \ref{tab:independentVariables} describes the metrics that we consider as independent variables: They belong to the three categories of metrics related to code review activity and engagement, \ie \textsl{`Participation'}, \textsl{`File coverage'}, and \textsl{`Activity'}. The variables in the \textsl{`Participation'} category are computed at review level (\eg they are equals among all the files in the same code review), while the variables in \textsl{`File coverage'} and \textsl{`Activity'} are computed at file level. 
                
                \smallskip
                \item [Statistical Modeling.] Before building our models, we ensure that our explanatory variables are not highly correlated with one another using Spearman rank correlation tests ($\rho$). We choose a rank correlation, instead of other types of correlation (\eg Pearson), because it is resilient to data that is not normally distributed. We consider a pair of variables highly correlated when $|\rho| > 0.7$, and only include one from the pair in the model, favoring the simplest.
            \end{description}
                
                Once we computed the variables of our study, we run \textsl{binary logistic regression} \cite{judge1982introduction} of the dependent variable:
                
                \begin{equation}
                    \begin{array}{l}
                    logit(\pi_{i,j}) = \beta_0 + \beta_1 \cdot reviewers_j +\\
                        \quad +  \beta_2 \cdot comments_{i,j} + \beta_3 \cdot words_{i,j} +\\
                        \quad +  ... (other~vars~and~\beta~omitted) + (1 | project)
                    \end{array}
                \end{equation}
                
                Here $logit(\pi_{i,j})$ represents the explained binary value (\textsl{`isDecreased'}) for a file $i$ in a review $j$, $\beta_0$ represents the log odds of being a code smell severity reduced in a review, while parameters $\beta_1 \cdot reviewers_j$, $\beta_2 \cdot comments_{i,j}$, $\beta_3 \cdot words_{i,j}$, etc. represent the differentials in the log odds of being a code smell severity reduced for a file reviewed in a review with characteristics $reviewers_{j-mean}$, $comments_{i,j-mean}$, $words_{i,j-mean}$, etc. Moreover, we run multilevel models to take into considerations that the reviewed files are nested into different projects (level 1), as neglecting a multilevel structure may lead to biased coefficients~\cite{robinson2009ecological}.                
                
                On the basis of the statistically significant codes given by the logistic regression model, we address \textbf{RQ$_1$} by verifying the significance of our independent variables, \ie the code review coverage, participation, and activity metrics.

        \subsection{\textbf{RQ$_2$}. Methodology}    
            In the second research question, we aim at understanding the influence of code review on specific types of code smells. We also investigate whether the types of code smell previously shown to be perceived as design problems by developers~\cite{palomba2014they,taibi2017developers} are treated differently. In the context of \textbf{RQ$_2$}, we exclude all the classes in which more than one code smell was detected. This is done to account for the observation-bias that might arise when performing analyses on classes affected by more smells at the same time: indeed, the perception of a specific type of smell can be biased if another design flaw co-occurs~\cite{abbes2011empirical,yamashita2013developers,palomba2018large}.
            
            We first report the number of code smells for each type (\eg `Complex Class') in our dataset: this helps us initially observe whether any trend in the observations exists or whether we should expect that the type of smell and the reported perception of the developers do not represent relevant factors for the reduction of the severity.
                        
            Subsequently, similarly to \textbf{RQ$_1$}, we verify if the statistical significance value given by the Chi-square test remains stable when controlling for additional factors. Therefore, we build a new binary regression model using the same dependent variable (\ie \textsl{`decreased'} or \textsl{`not decreased'}) as well as the same explanatory variables shown in Table \ref{tab:independentVariables}, to which we add a categorical variable representing the type of the smell (\eg `God Class'). Finally, we address \textbf{RQ$_2$} by verifying the significance of these variables: should they turn to be significant, it would indicate that the type of smell is related to the code smell severity reduction. We also analyze whether the significant smells (if any) are those reported as more important/perceived by developers when surveyed in previous literature.
            
        \subsection{\textbf{RQ$_3$}. Methodology}

            \input{tabs/tab-rq3}

            To answer \textbf{RQ$_3$}, we conduct a qualitative analysis of reviewers' discussions. We consider the code reviews in which we observe a variation in code smell severity (as done with \textbf{RQ$_2$}, we exclude code smell co-occurrences to avoid biases in our observations).
            Specifically, we perform an open card sort~\cite{nelson1976modified} that involves two of the authors of this paper (a graduate student and a research associate - who have more than seven years of programming experience each). Hereafter, we refer to them as the \emph{inspectors}. 
            An open card sort is a well-established technique used to extract salient themes and it allows the definition of taxonomies from input data \cite{nelson1976modified}. 
            In our case, first we used it to organize reviewers' discussions and classify whether the variation in code smell severity is due to (i) a specific request of the reviewers (\ie a reviewer explicitly asks the author to modify the code to reduce its smelliness) or (ii) a side-effect of the reviewers' requests (\ie no smell/design issue was mentioned, thus the smelliness was reduced as a consequence of other types of comments made by the reviewers). 
            
            In the cases in which the smelliness is reduced because of an explicit request of the reviewers, we further verify whether this is due to refactoring. 
            In remaining cases, we classify the side-effects leading to a variation of code smell severity, with the aim of understanding if there are particular reviewers' requests connected to the phenomenon of interest.
            To perform such an analysis, we extract a random statistically significant sample of \analyzedFiles cases\footnote{This corresponded to analyzing \analyzedReviews distinct reviews. Some reviews, in fact, contained more than one file with a decreased code smell severity.} in which the severity of a code smell in a file was reduced with a review from our dataset. Table~\ref{tab:rq3} reports the number of files and reviews inspected, by system.
            Afterwards, the process is conducted by two authors of this paper. They started by creating the list of \analyzedFiles cases to analyze. For each case, the list reported the file path and the link to the online code review. Then, each inspector independently analyzed half of the list. After inspecting the first few cases, the two inspectors opened a discussion on some unclear cases to reach a consensus on the names and types of both code smell symptoms and side-effects triggering a variation of the severity. In a few other cases, the other authors of the paper also participated in the discussion to validate the operations done by the two inspectors and suggest possible improvements.
            
        \subsection{Threats to Validity}            
        
        \begin{description}[leftmargin=0.5cm]
          \item[\textbf{Construct validity}.]
        
                We employed a heuristic-based code smell detector that computes code metrics and combines them to identify instances of the six considered design flaws. Such heuristics are inspired by those defined by Lanza and Marinescu~\cite{Lanza:2006}. After selecting the detection rules, we manually analyzed some code smell instances given as output of the detection and we applied more strict rules to reduce the number of false positives, which we deemed as too high. To allow replication of our study, the detector is publicly available in our online appendix~\cite{appendix}. 
                
                As a proxy to measure reviewers' perception of code smells (\textbf{RQ$_3$}), we considered whether a reviewer explicitly referred to a symptom of a smell within a discussion. To reduce the subjectivity of the classification, some cases where jointly evaluated by two inspectors.
                
                Finally, we relied on the reviewers' discussions to understand the reasons behind the variation of code smell severity. To try to mitigate the risk that developers may discuss design decisions through other channels~\cite{guzzi2013communication}, we select software systems whose developers are mostly remote~\cite{Bavota2015,mcintosh2014impact,Mcintosh2016} and that have a large number of code review data (thus indicating that developers actively use the code review platform). 
                
            \item[\textbf{Conclusion validity}.]
                To ensure that the selected binary logistic regression models (\textbf{RQ$_1$} and \textbf{RQ$_2$}) were appropriate for the data taken into account, we performed a number of preliminary checks such as: 
                (1) we ran a multilevel regression model~\cite{raudenbush2002hierarchical} to account for the multilevel structure determined by the presence of data from different projects; and (2) we built the models by adding the selected independent variables step-by-step and found that the coefficients remained stable, thus further indicating little to no interference among the variables.
                Furthermore, in our statistical model, we control for product and process metrics, which have been shown by previous research to be correlated to code smells~\cite{khomh2012exploratory,palomba2017diffuseness}.
                
            \item[\textbf{External validity}.]
                We conducted our study on the code review data belonging to a set of seven systems having different size and scope. Even though this size compares well with other experiments done in code review research~\cite{Bavota2015,mcintosh2014impact,Mcintosh2016,morales2015code}, a study investigating different projects may lead to different results. Furthermore, we only considered projects written in Java (due to the lack of techniques and tools able to accurately identify code smells in different programming languages \cite{de2018systematic,azeem2019machine}), thus results in other languages may vary. 
                
             \end{description}

%% file: tabs/tab-projects.tex
\begin{table}[!t]
	\caption{Dataset of our study.}
	\label{tab:projects}
	\resizebox{\columnwidth}{!}{%
		\begin{tabular}{lrrr}
		System & Time Span & Reviews & Reviewed Files \\
	
		\midrule
		\rowcolor{gray!30} couchbase-java-client & 01-12 to 11-17 & 916 & 4,479  \\
		couchbase-jvm-core & 04-14 to 11-17 & 841 & 4,217  \\
		\rowcolor{gray!30} eclipse.linuxtools & 06-12 to 11-17 & 4,129 & 23,088  \\
		eclipse.platform.ui & 02-13 to 11-17 & 4,756 & 52,441  \\
		\rowcolor{gray!30} egit & 10-09 to 11-17 & 5,336 & 18,647 \\
		jgit & 09-09 to 11-17 & 5,382 & 23,037  \\
		\rowcolor{gray!30} spymemcached & 05-10 to 07-17 & 519 & 2,777  \\
		
		\midrule
		 & \textit{Overall} & \textit{21,879} & \textit{128,691}  \\
		
		\end{tabular}
	}
\end{table}

%% file: tabs/tab-smells.tex

\begin{table*}[!t]
	\caption{Code smells considered in our study. The values in the column ``Perceived'' corresponds to: Y=Yes; N=No. }
	\label{tab:smells}
	\centering
	\resizebox{\textwidth}{!}{%
		\begin{tabular}{lp{9cm}p{2cm}}
	
		\toprule
		Code smell & Description & Perceived \\
	
		\midrule
		\rowcolor{gray!30} Complex Class \cite{brown1998antipatterns} & Classes exhibiting a high cyclomatic complexity, being therefore hard to test and understand \cite{abbes2011empirical,palomba2017diffuseness} & Y \cite{palomba2014they,taibi2017developers} \\
		
		Functional Decomposition \cite{fowler:1999} & Classes where object-oriented constructs are poorly used, declaring and implementing few methods \cite{fowler:1999}. & N \cite{taibi2017developers} \\
		
		\rowcolor{gray!30} God Class \cite{fowler:1999} & Classes that are poorly cohesive, usually characterized by large size and implementing several functionalities \cite{fowler:1999}. & Y \cite{palomba2014they,palomba2017scent} \\
		
		Inappropriate Intimacy \cite{fowler:1999} & Classes having a too high coupling with another class of the system \cite{fowler:1999}. & N \cite{palomba2014they,taibi2017developers} \\
		
		\rowcolor{gray!30} Lazy Class \cite{fowler:1999} & Classes generally composed of few lines of code and that have few relationships with other classes of the system \cite{fowler:1999}. According to Fowler \cite{fowler:1999}, this represents a problem because the time on maintaining these classes might be not worthy. & N \cite{palomba2014they} \\
		
		Spaghetti Code \cite{fowler:1999} & Classes without a structure that declare long methods without parameters \cite{fowler:1999}. & Y \cite{palomba2014they,taibi2017developers} \\
		
		\bottomrule
		\end{tabular}
	}
\end{table*}

%% file: tabs/tab-independentVariables.tex

\begin{table*}[!t]
	\caption{Independent and control variables used in our study.}
	\label{tab:independentVariables}
	\centering
	\resizebox{\textwidth}{!}{%
		\begin{tabular}{llp{7cm}p{7cm}}
		Type & Metric & Description & Motivation \\\hline\hline
		
		\multicolumn{4}{c}{Control variables}\\\hline
		\multirow{6}{*}{Product} & \cellcolor{gray!30}\multirow{2}{*}{Complexity$_{before,after}$} & \cellcolor{gray!30} The Weighted Method per Class metric computed on the version of the class before and after the review. & \cellcolor{gray!30} Classes with high complexity are potential candidates to be refactored. \\
		& \multirow{2}{*}{Size$_{before,after}$} & The Lines of Code metric computed on the version of the class before and after the review. & Large classes are hard to maintain and can be more prone to be subject of refactoring \cite{bavota2015experimental}. \\
		& \cellcolor{gray!30}\multirow{2}{*}{PatchSize} & \cellcolor{gray!30} Total number of files being subject of review. & \cellcolor{gray!30} Large patches can be more prone to be analyzed with respect to how the involved classes are designed. \\\hline
		
		\multirow{4}{*}{Process} &\multirow{2}{*}{Churn$_{before,after}$} & Sum of the lines added and removed in the version of the class before and after the review. & Classes having code smells are more change-prone \cite{khomh2012exploratory,palomba2017diffuseness}. \\
		& \cellcolor{gray!30}\multirow{2}{*}{PatchChurn} & \cellcolor{gray!30} Sum of the lines added or removed in all the classes being subject of review. & \cellcolor{gray!30} Large classes are hard to maintain and can be more prone to be subject of refactoring.\\\hline\hline
		
		\multicolumn{4}{c}{Independent variables}\\\hline
		\multirow{6}{*}{Participation} & \multirow{2}{*}{Reviewers} & \multirow{2}{*}{Number of reviewers involved in the code review.} & Classes having design issues may need more reviewers to be assessed. \\
		& \cellcolor{gray!30}\multirow{2}{*}{SelfApproved} & \cellcolor{gray!30} Whether the submitted patch is only approved for integration by the original author. & \cellcolor{gray!30} The original author already believes that the code is ready for integration, hence the patch has essentially not been reviewed.\\
		& \multirow{2}{*}{ReviewDuration} & \multirow{2}{*}{Time spent to perform the code review.} & Classes having design issues may enforce developers to spend more time discussing refactoring options.\\\hline
		
		\multirow{4}{*}{File coverage} & \cellcolor{gray!30}\multirow{2}{*}{PercentageComments} & \cellcolor{gray!30}Proportion of comments on the class under review to total number of comments. & \cellcolor{gray!30}Classes receiving less attention (such as comments) may display design issues. \\
		& \multirow{2}{*}{PercentageWords} & Proportion of words in the comments on the class to total number of words in all the comments. & Classes receiving less attention (such as number of words) may display design issues.\\\hline
		
		\multirow{4}{*}{Activity} & \cellcolor{gray!30}\multirow{2}{*}{Comments} & \cellcolor{gray!30}Number of comments made by reviewers on the class under review. & \cellcolor{gray!30}Classes having design issues can create more discussion among the reviewers on how to refactor them. \\
		& \multirow{2}{*}{Words} & Number of words in the comments made by reviewers on the class under review. & Classes having design issues can create deeper discussion among the reviewers on how to refactor them.\\

				
		\bottomrule
		\end{tabular}
	}
\end{table*}

%% file: tabs/tab-rq3.tex

\begin{table}[!t]
	\caption{Manually inspected cases in which the severity of a code smell in a file was reduced with a review, by system.}
	\label{tab:rq3}
	\resizebox{\columnwidth}{!}{%
		\begin{tabular}{lrr}
	
		System 										& Inspected Reviews & Inspected Files  \\
	
		\midrule
		\rowcolor{gray!30} couchbase-java-client 	& 33 				& 41  \\
		couchbase-jvm-core 							& 34 				& 47  \\
		\rowcolor{gray!30} eclipse.linuxtools 		& 33 				& 41  \\
		eclipse.platform.ui 						& 37 				& 69  \\
		\rowcolor{gray!30} egit 					& 59 				& 74 \\
		jgit 										& 52 				& 62  \\
		\rowcolor{gray!30} spymemcached 			& 24 				& 31  \\
		
		\midrule
		 \textit{Overall} 							& \textit{272} 	& \textit{365}  \\
		
		\end{tabular}
	}
\end{table}

%% file: sections/results.tex

    \begin{figure*}[ht]
            \centering
            \includegraphics[width=\textwidth]{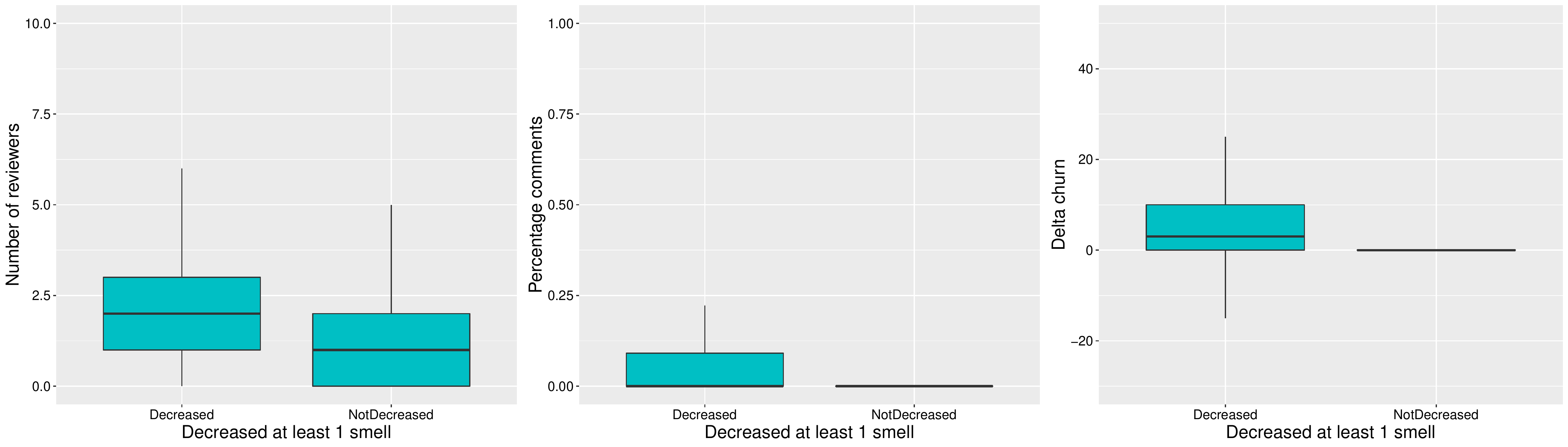}
            \caption{Number of reviewers, percentage of comments, and $\delta$-churn where code smell severity has been decreased and not.}
            \label{fig:bpAll}
            \vspace{-1em}
        \end{figure*}

\section{Analysis of the Results}
\label{sec:results}
    This section reports the results of the study, discussing each research question independently.

\subsection{$RQ_1$ The Influence Of Code Review On Code Smells}    
    Our analysis highlights that among 128,691 reviewed files (one file could be reviewed multiple times), 89,562 (70\%) were affected by at least one code smell before entering a review.
    The severity of the smells in these affected files is reduced only for a small fraction (4\%) with code reviews. This is in line with previous findings in the field~\cite{peters2012evaluating,chatzigeorgiou2010investigating,palomba2017scent}, which report how the degree of smelliness of classes tends to increase over time.

    Previous research showed that code review participation, percentage of discussion, and churn are good proxies to the goodness of a code review~\cite{spadini2018when,Thongtanunam2015,thongtanunam2016revisiting}. As a preliminary study, we compare these code review metrics and the decrease of severity using (i) the Wilcoxon rank sum test~\cite{conover} (with confidence level 95\%) and (ii) Cohen’s d~\cite{cliff} to estimate the magnitude of the observed difference. 
    We choose the Wilcoxon test since it is a non-parametric test (it does not have any assumption on the underlying data distribution), while we interpret the results of Cohen’s d relying on widely adopted guidelines, \ie negligible for $|\delta| < $ 0.10, small for 0.10 $ \leq |\delta| < $ 0.33, medium for 0.33 $ \leq |\delta| < $ 0.474, and large for $ |\delta| \geq $ 0.474~\cite{cliff}.

    Figure \ref{fig:bpAll} shows that code reviews with more participation, coverage, and activity are notably different with respect to the distribution of decreased and not decreased smell severity. In other words, the reviews with more reviewers, a higher percentage of comments, and more churn are those in which the severity of smells tends to decrease more often. In all the three cases, the differences observed in the distributions are statistically significant ($\alpha <$ 0.001), with a \emph{medium} effect size when considering the number of reviewers as well as $\delta$-churn, and a \emph{small} one in the case of the percentage of comments. Thus, on the basis of the results achieved so far, we see that (i) code smells generally do not decrease in terms of severity even if they are reviewed but (ii) the higher the quality of the code review process, the higher the likelihood to observe a reduction of code smell severity.
    
    \input{tabs/tab-model}
    
    We now consider \emph{all} the collected metrics (depicted in Table~\ref{tab:independentVariables}) using a regression model. From Table \ref{tab:modelResults} we observe that all the explanatory variables, including the ones referring to the code review process, appear to be statistically significant.\footnote{Some of the variables in Table \ref{tab:independentVariables} are not in the statistical models: These variables were removed after we found that they were collinear with others. In cases of collinearity, we kept the variable(s) that were simpler to compute and explain, in the interest of having a simpler model.} Thus, also controlling for possible confounding factors (such as the number of files reviewed and the size of the file under review), the three considered dimensions of code review represent a relevant factor to explain the reduction of code smell severity. We also evaluated the goodness of fit of our model, \ie how well the built model actually fits a set of observations. With multilevel logistic regression model, we cannot use the traditional \emph{Adjusted R$^2$} \cite{ohtani2000bootstrapping}, thus we used the method proposed by Nakagawa and Schielzeth~\cite{nakagawa2013general}: the \emph{Marginal R$^2$} (the variance explained by the fixed effects) measured 0.40 and the \emph{Conditional R$^2$} (the variance explained by the entire model) measured 0.45. These results are in line with the one reported by Morales \etal~\cite{morales2015code} when studying the impact of code review on design quality. 
    
    \begin{center}
        \begin{resultbox}
            \textbf{Finding 1:} In 96\% of the cases, the severity of the code smells in the files under review does not decrease. However, higher values of code review quality dimensions are indeed significantly related to a decrease in code smell severity.
        \end{resultbox}
    \end{center}

\subsection{$RQ_2$ The Influence Of Review On Smells, By Smell Type}    
    \input{tabs/tab-model_rq2_per_smell}
    
\begin{figure}[]
        \centering
        \includegraphics[width=\columnwidth]{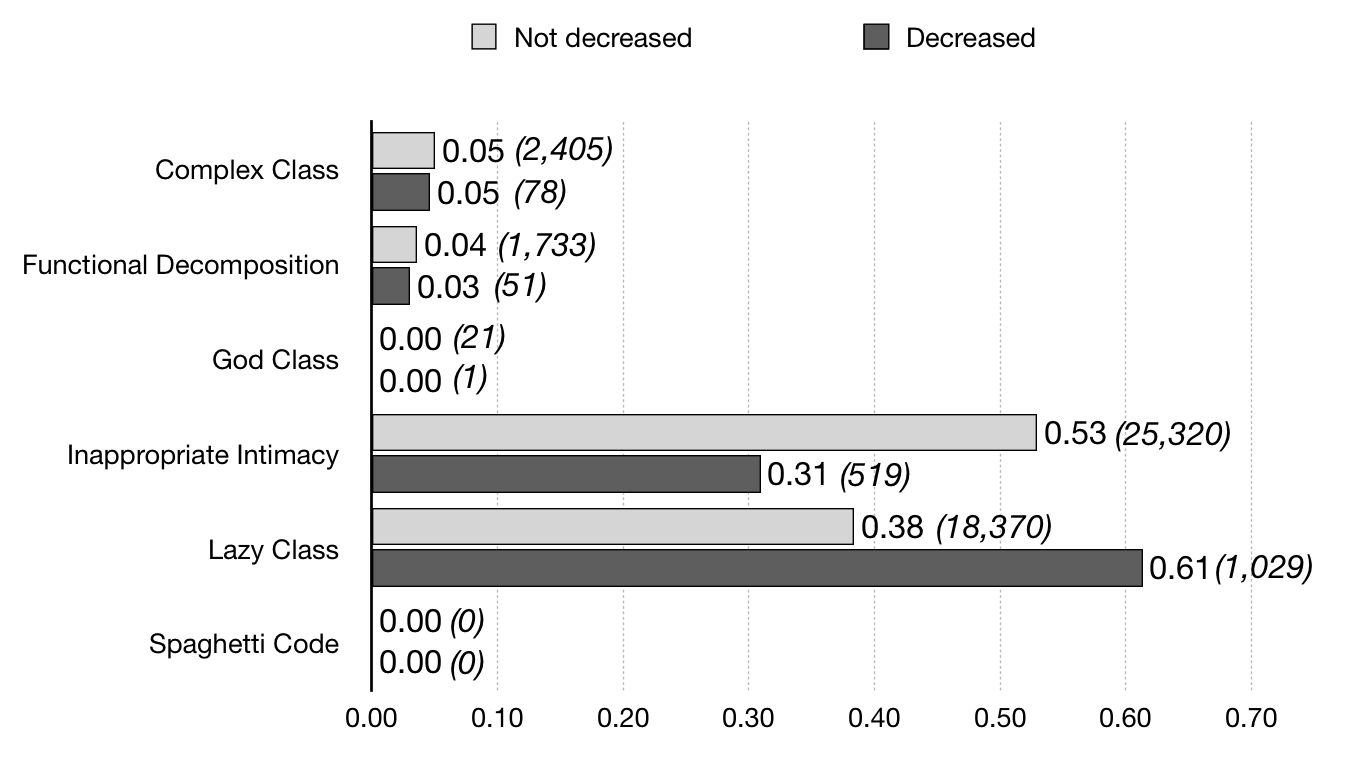}
        \caption{Decrease vs. non-decrease distribution, by smell type, excluding code smell co-occurrences.}
        \label{fig:rq21}
    \end{figure}

    Figure~\ref{fig:rq21} shows the distribution of decreased vs. not decreased severity, by type of smells. The Chi-square statistical test reports that the distributions differ significantly ($p << 0.01$). The biggest Chi-square contributions come from `Inappropriate Intimacy' and `Lazy Class' classes, while the others contribute almost nothing. Particularly, with an odds ratio~\cite{bland2000odds} of 2.5, `Lazy Class' is 2.5 times more likely to decrease after a code review than any other smell.
    
    Table~\ref{tab:modelrq2-per-smell} reports the results of the logistic regression modeling with the added independent variable `SmellType'. We also consider whether each smell was previously reported as perceived (P) or not perceived (NP) by developers in previous studies~\cite{palomba2014they,taibi2017developers,palomba2017scent}. Confirming the results of the aforementioned single-value analysis, `LazyClass' is statistically significant. Moreover, `God Class' is also significant, though with a lower value, which is most likely due to the lower overall incidence of this smell in the dataset (as visible in Figure~\ref{fig:rq21}). Furthermore, there is no trend on whether the smells that are expected to be ``more perceived'' by developers are those that are actually decreased more often in code review: In fact, both not perceived smells (\ie `LazyClass') and perceived ones (\ie `God Class') decrease with code review.

    \begin{center}
        \begin{resultbox}
            \textbf{Finding 2:} `Lazy Class' and `God Class' are the smells whose severity is the most likely to be reduced with a code review. We find no connection between previously reported developers' perceptions on types of code smells and the decrement of these smell types with reviews.
        \end{resultbox}
    \end{center}

\subsection{$RQ_3$ The Causes Of The Code Smell Severity Decrement}

    The findings of the previous \textbf{RQ}s showed that (i) code review participation, coverage, and activity are related to the reduction of code smell severity and (ii) it is the specific type of code smells rather than their actual perception to be relevant when explaining the reduction of severity. In this last research question we aim at shedding lights on (i) the actual motivations leading code smell severity to decrease during a code review and (ii) whether instances of the `Lazy Class', `Complex Class', and `God Class' smells, which appeared to be the most treated ones during code review in \textbf{RQ$_2$}, are subject to more refactoring activities than other smells.

    Table~\ref{tab:actions} reports the results of our manual analysis conducted on a statistically significant sample of \analyzedFiles cases in which the severity of a code smell in a file was reduced with a review. Our taxonomy is in line with that proposed by Tufano~\etal~\cite{tufano2017when}, who studied how code smells are removed over software evolution. We describe our findings by category.

    \input{tabs/tab-actions}

    \begin{description}[leftmargin=0.5cm]
        \item[\textsc{Code Addition}.] In this category, a discussion during a code review suggests improving the code by adding new statements. 
        Our manual analysis found 42\% of the code review discussions falling into this category.
        We could observe that code insertion requests are mainly concerned with the management of `Lazy Class' instances. 
        While the recommended operation in these cases is represented by the `Inline Class' refactoring (\ie the incorporation of the lazy class within an existing one \cite{fowler:1999}), we observed that in some cases reviewers recommend the addition of new functionalities, thus the insertion of new code. For example, this is the case for the class \texttt{CouchbaseClient} of the \textsc{couchbase-java-client} project: The reviewer explicitly requested the insertion of additional logging mechanisms that would have made the class more connected with the other classes of the system.

        \smallskip
        \item[\textsc{Code Replacement}.] The code review encourages the replacement of a certain piece of code and, consequently, the code smell (as well as the associated severity) can disappear, increase, or decrease. Typically, code review discussions in this category recommend a rewriting of the code from scratch instead of producing a refactoring of the original one. In our analysis, we observe that, in 32\% of the cases, the severity of code smells changes because of code replacements and it is a side effect rather than a direct action of the code review. Usually, reviewers suggest changes that correct defects or improve performance; following these suggestions, the author of the patch produces a new version with the effect of re-organizing the structure of the code and decrease the severity of the code smell.

        \smallskip
        \item[\textsc{Code Removal}.] The discussion during the code review leads to the suppression\footnote{Note that this category refers to code that is permanently removed.} of a piece of source code (\eg a few statements or a method). This removal affects artifacts connected to the code smell and the smell severity decreases. 
        Code review discussions that belong to this category often lead to a drop in the code smell severity as a side effect rather than a conscious decision. Our manual analysis confirms that the severity decreased due to indirect actions in 55 cases that, generically, refer to the suppression of `Blob' and `Complex Class' code smells. This is in line with past findings~\cite{tufano2017when}: sometimes the severity of code smells decreases as a side effect of code removal.

        \smallskip
        \item[\textsc{Refactoring}.] In this category, we refer to code changes that are explicitly replaced by applying one or multiple refactoring operations. The few refactoring actions requested within the considered reviewers' discussions (\ie 6\%) confirm the results of previous researchers~\cite{bavota2015experimental,murphy2012we,silva2016we,tufano2017when,tufano2016empirical}: Developers tend not to apply refactoring operations on classes affected by code smells. Here we see that such refactorings are not even recommended. However, in eight cases reviewers suggest action to reduce the code smell severity.

        \smallskip
        \item[\textsc{Major Restructuring}.] In this category, a reviewer recommends a restructuring of the system’s architecture that consequently changes the structure of the code under review. By nature, this category involves radical changes in the source code and, as such, it may implicitly include one of the above categories. However, it differs from the others since in this case we are not able to identify the exact code change leading to the smell removal. We could only see that it is a consequence of a major system’s restructuring. In our manual analysis we found that a major restructuring of the reviewed code is performed in 5\% of the cases and, as a side effect, this causes the smell severity to change.

        \smallskip
        \item[\textsc{Unclear}.] This category clusters the code changes that do not fit into the categories previously defined. In particular, this category contains all remaining code changes that reduce the associated code smells severity, but whose meaning or action is hard to understand.
        Only two cases from our manual analysis belong to this category. 
                    
    \end{description}

    \noindent
    Finally, the last column of Table~\ref{tab:actions} shows the number of times a code smell is referred to during code review. This number is very small: 19 times out of 365 (5\%). This result extends previous literature~\cite{Arcoverde:WRT2011,yamashita2013developers}\footnote{The authors found that developers avoid refactoring of code smells as it is not one of their top priorities~\cite{Arcoverde:WRT2011,yamashita2013developers}.} with the case of code review.

    Furthermore, during manual analysis, we searched for \emph{explicit references} to code smells, \ie ``We should refactor this class before it becomes a God class.''
    However, code smells are often symptoms of irregular trends of CK metrics (\eg code complexity, LOC, fan in or out, number of parameters) values. Hence, during code review, reviewers might discuss these problems (such as a complex class) without mentioning the code smell related to it (\ie `God Class'). In these cases, we did not consider these comments as ``referencing to a code smell'' because we wanted to investigate to what extent developers are aware specifically of code smells. 

    \begin{center}
        \begin{resultbox}
            \textbf{Finding 3:} In 95\% of the cases, the decrement in the severity of the code smells is a side effect of unrelated changes.
            The few cases in which reviewers explicitly suggest to reduce the code smell severity happen mostly as \textit{Code Addition} and \textit{Refactoring} changes.
        \end{resultbox}
    \end{center}

%% file: tabs/tab-model.tex

\begin{table}[h]
\caption{Multilevel logistic regression model.}
\centering
\label{tab:modelResults}
\resizebox{\columnwidth}{!}{%

\begin{tabular}{cccc}
\multicolumn{1}{l|}{}                                             & \multicolumn{3}{c|}{\cellcolor[HTML]{C0C0C0}\textbf{Has the severity of any smell decreased?}}                                                          \\
\multicolumn{1}{l|}{}                                             & \cellcolor[HTML]{C0C0C0}Estimate & \cellcolor[HTML]{C0C0C0}S.E. & \multicolumn{1}{l|}{\cellcolor[HTML]{C0C0C0}Significance} \\\hline

\multicolumn{1}{l|}{\cellcolor[HTML]{EFEFEF}Intercept}            & \multicolumn{1}{r}{-2.762e+00}  & \multicolumn{1}{r}{2.031e-01}  & \multicolumn{1}{l|}{***}                          \\
\multicolumn{1}{l|}{\cellcolor[HTML]{EFEFEF}PatchSize}            & \multicolumn{1}{r}{-9.360e-04}  & \multicolumn{1}{r}{8.489e-05}  & \multicolumn{1}{l|}{***}                          \\
\multicolumn{1}{l|}{\cellcolor[HTML]{EFEFEF}ReviewDuration}       & \multicolumn{1}{r}{7.676e-05}   & \multicolumn{1}{r}{6.563e-06}  & \multicolumn{1}{l|}{***}                          \\
\multicolumn{1}{l|}{\cellcolor[HTML]{EFEFEF}SizeBefore}           & \multicolumn{1}{r}{1.111e-04}   & \multicolumn{1}{r}{2.764e-05}  & \multicolumn{1}{l|}{***}                           \\
\multicolumn{1}{l|}{\cellcolor[HTML]{EFEFEF}ChurnBefore}          & \multicolumn{1}{r}{2.490e-04}   & \multicolumn{1}{r}{8.812e-05}  & \multicolumn{1}{l|}{**}                           \\
\multicolumn{1}{l|}{\cellcolor[HTML]{EFEFEF}DeltaChurn}           & \multicolumn{1}{r}{1.056e-03}   & \multicolumn{1}{r}{4.091e-04}  & \multicolumn{1}{l|}{**}                           \\
\multicolumn{1}{l|}{\cellcolor[HTML]{EFEFEF}Comments}             & \multicolumn{1}{r}{1.476e-02}   & \multicolumn{1}{r}{6.522e-04}  & \multicolumn{1}{l|}{***}                          \\
\multicolumn{1}{l|}{\cellcolor[HTML]{EFEFEF}PercentageComments}   & \multicolumn{1}{r}{1.736e+00}   & \multicolumn{1}{r}{6.716e-02}  & \multicolumn{1}{l|}{***}                          \\
\multicolumn{1}{l|}{\cellcolor[HTML]{EFEFEF}SelfApproved}         & \multicolumn{1}{r}{-1.007e+00}  & \multicolumn{1}{r}{6.335e-02}  & \multicolumn{1}{l|}{***}                          \\\hline\hline
\multicolumn{4}{c}{marginal R$^2$ = 0.40, conditional R$^2$ = 0.45}
\\\hline\hline
\multicolumn{4}{r}{\emph{Significance codes: '***'$p<$0.001, '**'$p<$0.01, '*'$p<$0.05, '.'$p<$0.1}}
\\                                                                                                                                                                                                            
\end{tabular}

}
\end{table}


%% file: tabs/tab-model_rq2_per_smell.tex

\begin{table}[]
\caption{Multilevel logistic regression model with the additional independent variable \textit{SmellType}. P and NP specify whether the specific smell was reported as being perceived by developers in previous studies~\cite{palomba2014they,taibi2017developers,palomba2017scent}.}
\centering
\label{tab:modelrq2-per-smell}
\resizebox{\columnwidth}{!}{%

\begin{tabular}{cccc}
\multicolumn{1}{l|}{}                                                                               & \multicolumn{3}{c|}{\cellcolor[HTML]{C0C0C0}\textbf{Has the severity of the smell decreased?}}                                                          \\
\multicolumn{1}{l|}{}                                                                               & \cellcolor[HTML]{C0C0C0}Estimate & \cellcolor[HTML]{C0C0C0}S.E. & \multicolumn{1}{l|}{\cellcolor[HTML]{C0C0C0}Significance} \\\hline

\multicolumn{1}{l|}{\cellcolor[HTML]{EFEFEF}Intercept}                                          & \multicolumn{1}{r}{-2.889e+00}   & \multicolumn{1}{r}{2.387e-01}   & \multicolumn{1}{l|}{***}                          \\
\multicolumn{1}{l|}{\cellcolor[HTML]{EFEFEF}PatchSize}                                          & \multicolumn{1}{r}{-1.501e-03}   & \multicolumn{1}{r}{1.466e-04}   &  \multicolumn{1}{l|}{***}                          \\
\multicolumn{1}{l|}{\cellcolor[HTML]{EFEFEF}ReviewDuration}                                     & \multicolumn{1}{r}{7.782e-05}    & \multicolumn{1}{r}{9.179e-06}   &  \multicolumn{1}{l|}{***}                          \\
\multicolumn{1}{l|}{\cellcolor[HTML]{EFEFEF}SizeBefore}                                         & \multicolumn{1}{r}{-2.721e-03}   & \multicolumn{1}{r}{2.646e-04}   &  \multicolumn{1}{l|}{***}                          \\
\multicolumn{1}{l|}{\cellcolor[HTML]{EFEFEF}ChurnBefore}                                        & \multicolumn{1}{r}{1.615e-03}    & \multicolumn{1}{r}{2.978e-04}   &  \multicolumn{1}{l|}{***}                          \\
\multicolumn{1}{l|}{\cellcolor[HTML]{EFEFEF}DeltaChurn}                                         & \multicolumn{1}{r}{5.765e-03}    & \multicolumn{1}{r}{9.813e-04}   &  \multicolumn{1}{l|}{***}                          \\
\multicolumn{1}{l|}{\cellcolor[HTML]{EFEFEF}Comments}                                        	& \multicolumn{1}{r}{1.340e-02}    & \multicolumn{1}{r}{8.019e-04}   &  \multicolumn{1}{l|}{***}                          \\
\multicolumn{1}{l|}{\cellcolor[HTML]{EFEFEF}PercentageComments}                                 & \multicolumn{1}{r}{1.898e+00}    & \multicolumn{1}{r}{1.155e-01}   &  \multicolumn{1}{l|}{***}                          \\
\multicolumn{1}{l|}{\cellcolor[HTML]{EFEFEF}SelfApproved}                                 		& \multicolumn{1}{r}{-1.326e+00}   & \multicolumn{1}{r}{9.992e-02}   &  \multicolumn{1}{l|}{***}                          \\
\multicolumn{1}{l|}{\cellcolor[HTML]{EFEFEF}(P) ComplexClass}                 					& \multicolumn{1}{r}{3.724e-01}    & \multicolumn{1}{r}{2.259e-01}   &  \multicolumn{1}{l|}{.}                             \\
\multicolumn{1}{l|}{\cellcolor[HTML]{EFEFEF}(P) GodClass}                     					& \multicolumn{1}{r}{3.025e+00}    & \multicolumn{1}{r}{1.224e+00}   &  \multicolumn{1}{l|}{*}                             \\
\multicolumn{1}{l|}{\cellcolor[HTML]{EFEFEF}(NP) InappropriateIntimacy}        					& \multicolumn{1}{r}{-1.545e-01}   & \multicolumn{1}{r}{1.699e-01}   &  \multicolumn{1}{l|}{}                            \\
\multicolumn{1}{l|}{\cellcolor[HTML]{EFEFEF}(NP) LazyClass}                    					& \multicolumn{1}{r}{6.949e-01}    & \multicolumn{1}{r}{1.678e-01}   &  \multicolumn{1}{l|}{***}                          \\\hline\hline
\multicolumn{4}{c}{marginal R$^2$ = 0.62, conditional R$^2$ = 0.64}
\\\hline\hline
\multicolumn{4}{r}{\emph{Significance codes: '***'$p<$0.001, '**'$p<$0.01, '*'$p<$0.05, '.'$p<$0.1}}
\\

\end{tabular}
}
\end{table}


%% file: tabs/tab-actions.tex

\begin{table}[h]
\centering
	\caption{Code smells categories manually inspected. }
	\label{tab:actions}

\begin{tabular}{l|rr|r}
Category            & Count & Percentage (\%) 	& Ref. to smells 	\\ \hline
\rowcolor[HTML]{EFEFEF} 
Code Addition      & 152 	& 41.6\%            & 6   				\\
Code Removal        & 55 	& 15.1\% 			& 0               	\\
\rowcolor[HTML]{EFEFEF} 
Code Replacement    & 116 	& 31.8\%            & 2  				\\
Refactoring         & 23 	& 6.3\%             & 8	 				\\
\rowcolor[HTML]{EFEFEF} 
Major Restructuring & 17 	& 4.7\%             & 3 				\\
Unclear             & 2 	& 0.5\%             & 0  				\\ \hline
\rowcolor[HTML]{EFEFEF} 
\textit{Overall}         	& \textit{365} 	& \textit{100\%}            & \textit{19} 				\\
\end{tabular}
\end{table}

%% file: sections/discussions.tex
\section{Discussion and Implications}
	Our results highlighted some points to be further discussed as well as a set of practical implications for both the research community and tool vendors. In particular:
	
	\begin{description}[leftmargin=0.5cm]
		\item[Keep the quality of code review high.] We find that code reviews in which participation, coverage, and activity are higher tend to lead to a non-targeted, yet significant reduction in code smell severity. As an outcome, we confirm the importance of the research effort devoted to the definition of methodologies and novel approaches helping practitioners keep the quality of the review process high~\cite{sadowski2018modern,thongtanunam2015should}. It is, thus, important when thinking of code review as a set of activities that, in addition to finding defects, include knowledge transfer or team awareness~\cite{bacchelli:2013}. In these conditions, the presence of smells in the code might produce improper learning or misleading assumptions.
		Consequently, our study, in line with Pascarella \etal~\cite{pascarella2018information}, revealed the need for further mechanisms to stimulate developers and make them aware of the advantages of participating and actively discussing in code review. These observations impact on educational aspects, which researchers in the code review field should further investigate and promote, but also to the definition of effective involvement strategies (\eg code review gamification~\cite{unkelos2015gamifying}), which are still under-investigated by the research community.\smallskip
	
		\item[On the developers' awareness of code smells.] A clear outcome of our research is that the severity of code smells is often left intact or even increased during code review. This result may be a consequence that developers do not perceive code smells, that code smells cannot be properly visualized, or even that developers prefer to ignore these design issues. Moreover, our findings confirmed a problem that the research community already pointed out in the past: developers tend not to refactor or deal with code smells~\cite{bavota2015experimental,silva2016we,palomba2017exploratory,kim2014empirical} and, unfortunately, code review does not seem to change the state of things either. We believe that this outcome has implications for a broad set of researchers. First, effective and contextual code smell detectors and static analyzers~\cite{pantiuchina2018towards,sae2016context,vassallo2018context}, able to pinpoint the presence of design issues within the scope of a commit, should be made available to reviewers; as such, researchers in the field of code smells are called to devise new methods and tools allowing a just-in-time detection that can be applied at code review time. Second, we point out a problem of visualization of code smell-related information in code review: indeed, one possible reason behind the results achieved in our research is that existing code review tools are too focused on the changed lines and do not offer an overview of the classes under analysis. In other words, code review tools are yet unable to signal to reviewers the presence of design flaws. Thus, an implication of our study is that augmenting code review tools with new visualizations that support the detection and correction of code smells could be beneficial. Finally, our findings support the research on how to order code changes within code review~\cite{baum2017optimal,spadini2019test}, which may be used as a way to prioritize the analysis and correction of code smells.\smallskip
		
		\item[Promoting smart discussions in code review.] As both high participation and discussion within the code review process tend to promote the application of changes that have the side-effect of reducing code smell severity, a possible outcome of our research consists in calling for the definition of novel methodologies that directly address the people involved in the code review rather than the process itself. For example, novel machine learning-based methods that recognize the current context of a discussion and propose smart suggestions to lead reviewers discussing of specific design flaws that impact the maintainability of source code might be a potentially useful way to foster code quality within the code review process. Similarly, the recent advances on code documentation \cite{pascarella2017classifying,pascarella2019classifying,sridhara2010towards}, showing and classifying the comments that help developers understanding source code, might be put in the context of code review to provide developers with further contextual information to locate technical debts.\smallskip
		
	\end{description}

%% file: sections/conclusions.tex

\section{Conclusions}
\label{sec:conclusions}
	In the study presented in this paper, we investigated the relation of code review to code smells by mining more than 21,000 code reviews belonging to seven Java open-source projects and considering six well-known code smells.
	Our findings pointed out that active and participated code reviews have a significant influence on the reduction of code smell severity. However, such a reduction is typically a side effect of code changes that are not necessarily related to code smells. 
	The results of the study represent the input for our future research agenda, which primarily includes the definition and evaluation of non-intrusive alert systems that notify the presence of code smells at code review-time. Furthermore, we aim at replicating our study considering projects from different domains as well as closed-source projects, which may have different guidelines for code review activities.

%% file: saner_main.bbl
\begin{thebibliography}{10}

\bibitem{fowler:1999}
M.~Fowler, K.~Beck, J.~Brant, W.~Opdyke, and D.~Roberts, {\em Refactoring:
  Improving the Design of Existing Code}.
\newblock Addison-Wesley, 1999.

\bibitem{palomba2017diffuseness}
F.~Palomba, G.~Bavota, M.~Di~Penta, F.~Fasano, R.~Oliveto, and A.~De~Lucia,
  ``On the diffuseness and the impact on maintainability of code smells: a
  large scale empirical investigation,'' {\em Empirical Software Engineering},
  pp.~1--34, 2017.

\bibitem{olbrich2009evolution}
S.~Olbrich, D.~S. Cruzes, V.~Basili, and N.~Zazworka, ``The evolution and
  impact of code smells: A case study of two open source systems,'' in {\em
  Empirical Software Engineering and Measurement, 2009. ESEM 2009. 3rd
  International Symposium on}, pp.~390--400, IEEE, 2009.

\bibitem{khomh2012exploratory}
F.~Khomh, M.~Di~Penta, Y.-G. Gu{\'e}h{\'e}neuc, and G.~Antoniol, ``An
  exploratory study of the impact of antipatterns on class change-and
  fault-proneness,'' {\em Empirical Software Engineering}, vol.~17, no.~3,
  pp.~243--275, 2012.

\bibitem{Spadini:icsme18}
D.~Spadini, F.~Palomba, A.~Zaidman, M.~Bruntink, and A.~Bacchelli, ``On the
  relation of test smells to software code quality,'' in {\em 2018 IEEE
  International Conference on Software Maintenance and Evolution (ICSME)},
  pp.~1--12, IEEE, 2018.

\bibitem{abbes2011empirical}
M.~Abbes, F.~Khomh, Y.-G. Gueheneuc, and G.~Antoniol, ``An empirical study of
  the impact of two antipatterns, blob and spaghetti code, on program
  comprehension,'' in {\em Software maintenance and reengineering (CSMR), 2011
  15th European conference on}, pp.~181--190, IEEE, 2011.

\bibitem{sjoberg2013quantifying}
D.~I. Sjoberg, A.~Yamashita, B.~C. Anda, A.~Mockus, and T.~Dyba, ``Quantifying
  the effect of code smells on maintenance effort,'' {\em IEEE Transactions on
  Software Engineering}, no.~8, pp.~1144--1156, 2013.

\bibitem{tufano2017when}
M.~Tufano, F.~Palomba, G.~Bavota, R.~Oliveto, M.~Di~Penta, A.~De~Lucia, and
  D.~Poshyvanyk, ``When and why your code starts to smell bad (and whether the
  smells go away),'' {\em IEEE Transactions on Software Engineering}, vol.~43,
  no.~11, pp.~1063--1088, 2017.

\bibitem{tufano2016empirical}
M.~Tufano, F.~Palomba, G.~Bavota, M.~Di~Penta, R.~Oliveto, A.~De~Lucia, and
  D.~Poshyvanyk, ``An empirical investigation into the nature of test smells,''
  in {\em Proceedings of the 31st IEEE/ACM International Conference on
  Automated Software Engineering}, pp.~4--15, 2016.

\bibitem{peters2012evaluating}
R.~Peters and A.~Zaidman, ``Evaluating the lifespan of code smells using
  software repository mining,'' in {\em Software Maintenance and Reengineering
  (CSMR), 2012 16th European Conference on}, pp.~411--416, IEEE, 2012.

\bibitem{bavota2015experimental}
G.~Bavota, A.~De~Lucia, M.~Di~Penta, R.~Oliveto, and F.~Palomba, ``An
  experimental investigation on the innate relationship between quality and
  refactoring,'' {\em Journal of Systems and Software}, vol.~107, pp.~1--14,
  2015.

\bibitem{palomba2017scent}
F.~Palomba, A.~Panichella, A.~Zaidman, R.~Oliveto, and A.~De~Lucia, ``The scent
  of a smell: An extensive comparison between textual and structural smells,''
  {\em Transactions on Software Engineering}, vol.~44, no.~10, pp.~977--1000,
  2018.

\bibitem{palomba2018beyond}
F.~Palomba, D.~A.~A. Tamburri, F.~A. Fontana, R.~Oliveto, A.~Zaidman, and
  A.~Serebrenik, ``Beyond technical aspects: How do community smells influence
  the intensity of code smells?,'' {\em IEEE transactions on software
  engineering}, 2018.

\bibitem{YamashitaM12}
A.~F. Yamashita and L.~Moonen, ``Do code smells reflect important
  maintainability aspects?,'' in {\em Proceedings of the International
  Conference on Software Maintenance (ICSM)}, pp.~306--315, IEEE, 2012.

\bibitem{yamashita2013developers}
A.~Yamashita and L.~Moonen, ``Do developers care about code smells? an
  exploratory survey,'' in {\em Reverse Engineering (WCRE), 2013 20th Working
  Conference on}, pp.~242--251, IEEE, 2013.

\bibitem{palomba2014they}
F.~Palomba, G.~Bavota, M.~Di~Penta, R.~Oliveto, and A.~De~Lucia, ``Do they
  really smell bad? a study on developers' perception of bad code smells,'' in
  {\em Software maintenance and evolution (ICSME), 2014 IEEE international
  conference on}, pp.~101--110, IEEE, 2014.

\bibitem{taibi2017developers}
D.~Taibi, A.~Janes, and V.~Lenarduzzi, ``How developers perceive smells in
  source code: A replicated study,'' {\em Information and Software Technology},
  vol.~92, pp.~223--235, 2017.

\bibitem{silva2016we}
D.~Silva, N.~Tsantalis, and M.~T. Valente, ``Why we refactor? confessions of
  github contributors,'' in {\em Proceedings of the 2016 24th ACM SIGSOFT
  International Symposium on Foundations of Software Engineering},
  pp.~858--870, ACM, 2016.

\bibitem{palomba2017exploratory}
F.~Palomba, A.~Zaidman, R.~Oliveto, and A.~De~Lucia, ``An exploratory study on
  the relationship between changes and refactoring,'' in {\em Program
  Comprehension (ICPC), 2017 IEEE/ACM 25th International Conference on},
  pp.~176--185, IEEE, 2017.

\bibitem{kim2014empirical}
M.~Kim, T.~Zimmermann, and N.~Nagappan, ``An empirical study of refactoring
  challenges and benefits at microsoft,'' {\em IEEE Transactions on Software
  Engineering}, vol.~40, no.~7, pp.~633--649, 2014.

\bibitem{morales2015code}
R.~Morales, S.~McIntosh, and F.~Khomh, ``Do code review practices impact design
  quality? a case study of the qt, vtk, and itk projects,'' in {\em Software
  Analysis, Evolution and Reengineering (SANER), 2015 IEEE 22nd International
  Conference on}, pp.~171--180, IEEE, 2015.

\bibitem{bacchelli:2013}
A.~Bacchelli and C.~Bird, ``Expectations, outcomes, and challenges of modern
  code review,'' in {\em 35th International Conference on Software
  Engineering}, pp.~712--721, IEEE, 2013.

\bibitem{appendix}
L.~Pascarella, D.~Spadini, F.~Palomba, and A.~Bacchelli, ``Data and materials
  \url{\datamaterial},'' 2019.

\bibitem{Abelein2015}
U.~Abelein and B.~Paech, ``{Understanding the Influence of User Participation
  and Involvement on System Success: a Systematic Mapping Study},'' {\em
  Empirical Software Engineering}, vol.~20, no.~1, pp.~28--81, 2015.

\bibitem{Thongtanunam2015}
P.~Thongtanunam, S.~McIntosh, A.~E. Hassan, and H.~Iida, ``Investigating code
  review practices in defective files: An empirical study of the qt system,''
  in {\em MSR '15 Proceedings of the 12th Working Conference on Mining Software
  Repositories}, 2015.

\bibitem{thongtanunam2016revisiting}
P.~Thongtanunam, S.~McIntosh, A.~E. Hassan, and H.~Iida, ``Revisiting code
  ownership and its relationship with software quality in the scope of modern
  code review,'' in {\em Proceedings of the 38th international conference on
  software engineering}, pp.~1039--1050, ACM, 2016.

\bibitem{Rigby2014}
P.~C. Rigby, D.~M. German, L.~Cowen, and M.-A. Storey, ``Peer review on open
  source software projects: Parameters, statistical models, and theory,'' {\em
  ACM Transactions on Software Engineering and Methodology}, p.~34, 2014.

\bibitem{porter1998comparing}
A.~Porter and L.~Votta, ``Comparing detection methods for software requirements
  inspections: A replication using professional subjects,'' {\em Empirical
  software engineering}, vol.~3, no.~4, pp.~355--379, 1998.

\bibitem{Sauer2000}
C.~Sauer, D.~R. Jeffery, L.~Land, and P.~Yetton, ``The effectiveness of
  software development technical reviews: A behaviorally motivated program of
  research,'' {\em Software Engineering, IEEE Transactions on}, vol.~26, no.~1,
  pp.~1--14, 2000.

\bibitem{kemerer2009impact}
C.~F. Kemerer and M.~C. Paulk, ``The impact of design and code reviews on
  software quality: An empirical study based on psp data,'' {\em IEEE
  transactions on software engineering}, vol.~35, no.~4, pp.~534--550, 2009.

\bibitem{Rigby2013}
P.~C. Rigby and C.~Bird, ``Convergent contemporary software peer review
  practices,'' in {\em Proceedings of the 2013 9th Joint Meeting on Foundations
  of Software Engineering}, pp.~202--212, ACM, 2013.

\bibitem{kononenko2015investigating}
O.~Kononenko, O.~Baysal, L.~Guerrouj, Y.~Cao, and M.~W. Godfrey,
  ``Investigating code review quality: Do people and participation matter?,''
  in {\em 2015 IEEE international conference on software maintenance and
  evolution (ICSME)}, pp.~111--120, IEEE, 2015.

\bibitem{di2016security}
M.~di~Biase, M.~Bruntink, and A.~Bacchelli, ``A security perspective on code
  review: The case of chromium,'' in {\em 2016 IEEE 16th International Working
  Conference on Source Code Analysis and Manipulation (SCAM)}, pp.~21--30,
  IEEE, 2016.

\bibitem{Thongtanunam2016}
P.~Thongtanunam, S.~Mcintosh, A.~E. Hassan, and H.~Iida, ``{Review
  Participation in Modern Code Review},'' {\em Empirical Software Engineering
  (EMSE)}, vol.~22, no.~2, pp.~768--817, 2017.

\bibitem{kononenko2016code}
O.~Kononenko, O.~Baysal, and M.~W. Godfrey, ``Code review quality: how
  developers see it,'' in {\em 2016 IEEE/ACM 38th International Conference on
  Software Engineering (ICSE)}, pp.~1028--1038, IEEE, 2016.

\bibitem{Bavota2015}
G.~Bavota and B.~Russo, ``{Four eyes are better than two: On the impact of code
  reviews on software quality},'' in {\em 2015 IEEE 31st International
  Conference on Software Maintenance and Evolution, ICSME 2015 - Proceedings},
  pp.~81--90, 2015.

\bibitem{mcintosh2014impact}
S.~McIntosh, Y.~Kamei, B.~Adams, and A.~E. Hassan, ``The impact of code review
  coverage and code review participation on software quality: A case study of
  the qt, vtk, and itk projects,'' in {\em Proceedings of the 11th Working
  Conference on Mining Software Repositories}, pp.~192--201, ACM, 2014.

\bibitem{Mcintosh2016}
S.~McIntosh, Y.~Kamei, B.~Adams, and A.~E. Hassan, ``{An empirical study of the
  impact of modern code review practices on software quality},'' {\em Empirical
  Software Engineering}, vol.~21, no.~5, pp.~2146--2189, 2016.

\bibitem{sadowski2018modern}
C.~Sadowski, E.~S{\"o}derberg, L.~Church, M.~Sipko, and A.~Bacchelli, ``Modern
  code review: a case study at google,'' in {\em Proceedings of the 40th
  International Conference on Software Engineering: Software Engineering in
  Practice}, pp.~181--190, ACM, 2018.

\bibitem{ram2018makes}
A.~Ram, A.~A. Sawant, M.~Castelluccio, and A.~Bacchelli, ``What makes a code
  change easier to review: an empirical investigation on code change
  reviewability,'' in {\em Proceedings of the 2018 26th ACM Joint Meeting on
  European Software Engineering Conference and Symposium on the Foundations of
  Software Engineering}, pp.~201--212, ACM, 2018.

\bibitem{pascarella2018information}
L.~Pascarella, D.~Spadini, F.~Palomba, M.~Bruntink, and A.~Bacchelli,
  ``Information needs in contemporary code review,'' {\em Proceedings of the
  ACM on Human-Computer Interaction}, vol.~2, no.~CSCW, p.~135, 2018.

\bibitem{mantyla2009types}
M.~V. M{\"a}ntyl{\"a} and C.~Lassenius, ``What types of defects are really
  discovered in code reviews?,'' {\em IEEE Transactions on Software
  Engineering}, vol.~35, no.~3, pp.~430--448, 2009.

\bibitem{beller2014modern}
M.~Beller, A.~Bacchelli, A.~Zaidman, and E.~Juergens, ``Modern code reviews in
  open-source projects: Which problems do they fix?,'' in {\em Proceedings of
  the 11th working conference on mining software repositories}, pp.~202--211,
  ACM, 2014.

\bibitem{Arcoverde:WRT2011}
R.~Arcoverde, A.~Garcia, and E.~Figueiredo, ``Understanding the longevity of
  code smells: preliminary results of an explanatory survey,'' in {\em
  International Workshop on Refactoring Tools}, pp.~33--36, ACM, 2011.

\bibitem{rahman2013and}
F.~Rahman and P.~Devanbu, ``How, and why, process metrics are better,'' in {\em
  2013 35th International Conference on Software Engineering (ICSE)},
  pp.~432--441, IEEE, 2013.

\bibitem{Paixao2018}
M.~Paixao, J.~Krinke, D.~Han, and M.~Harman, ``Crop: Linking code reviews to
  source code changes,'' in {\em International Conference on Mining Software
  Repositories}, MSR, 2018.

\bibitem{voigt2017eu}
P.~Voigt and A.~Von~dem Bussche, ``The eu general data protection regulation
  (gdpr),'' {\em A Practical Guide, 1st Ed., Cham: Springer International
  Publishing}, 2017.

\bibitem{Lanza:2006}
M.~Lanza and R.~Marinescu, {\em Object-Oriented Metrics in Practice: Using
  Software Metrics to Characterize, Evaluate, and Improve the Design of
  Object-Oriented Systems}.
\newblock Springer, 2006.

\bibitem{brown1998antipatterns}
W.~H. Brown, R.~C. Malveau, H.~W. McCormick, and T.~J. Mowbray, {\em
  AntiPatterns: refactoring software, architectures, and projects in crisis}.
\newblock John Wiley \& Sons, Inc., 1998.

\bibitem{saaty1993relative}
T.~L. Saaty, ``What is relative measurement? the ratio scale phantom,'' {\em
  Mathematical and Computer Modelling}, vol.~17, no.~4-5, pp.~1--12, 1993.

\bibitem{pecorelli2019comparing}
F.~Pecorelli, F.~Palomba, D.~Di~Nucci, and A.~De~Lucia, ``Comparing heuristic
  and machine learning approaches for metric-based code smell detection,'' in
  {\em Proceedings of the 27th International Conference on Program
  Comprehension}, pp.~93--104, IEEE Press, 2019.

\bibitem{CKMetrics}
S.~Chidamber and C.~Kemerer, ``A metrics suite for object oriented design,''
  {\em IEEE Trans. on Software Engineering}, vol.~20, no.~6, pp.~476--493,
  1994.

\bibitem{fernandes2016review}
E.~Fernandes, J.~Oliveira, G.~Vale, T.~Paiva, and E.~Figueiredo, ``A
  review-based comparative study of bad smell detection tools,'' in {\em
  Proceedings of the 20th International Conference on Evaluation and Assessment
  in Software Engineering}, p.~18, ACM, 2016.

\bibitem{garcia2016improved}
J.~Garc{\'\i}a-Munoz, M.~Garc{\'\i}a-Valls, and J.~Escribano-Barreno,
  ``Improved metrics handling in sonarqube for software quality monitoring,''
  in {\em Distributed Computing and Artificial Intelligence, 13th International
  Conference}, pp.~463--470, Springer, 2016.

\bibitem{ferenc2014source}
R.~Ferenc, L.~Lang{\'o}, I.~Siket, T.~Gyim{\'o}thy, and T.~Bakota, ``Source
  meter sonar qube plug-in,'' in {\em Source Code Analysis and Manipulation
  (SCAM), 2014 IEEE 14th International Working Conference on}, pp.~77--82,
  IEEE, 2014.

\bibitem{lopez2005relevance}
M.~Lopez and N.~Habra, ``Relevance of the cyclomatic complexity threshold for
  the java programming language,'' {\em SMEF 2005}, p.~195, 2005.

\bibitem{fontana2015towards}
F.~A. Fontana, V.~Ferme, M.~Zanoni, and R.~Roveda, ``Towards a prioritization
  of code debt: A code smell intensity index,'' in {\em 2015 IEEE 7th
  International Workshop on Managing Technical Debt (MTD)}, pp.~16--24, IEEE,
  2015.

\bibitem{judge1982introduction}
G.~G. Judge, R.~C. Hill, W.~Griffiths, H.~Lutkepohl, and T.~C. Lee, {\em
  Introduction to the Theory and Practice of Econometrics.}
\newblock New York New York John Wiley and Sons 1982., 1982.

\bibitem{robinson2009ecological}
W.~S. Robinson, ``Ecological correlations and the behavior of individuals,''
  {\em International journal of epidemiology}, vol.~38, no.~2, pp.~337--341,
  2009.

\bibitem{palomba2018large}
F.~Palomba, G.~Bavota, M.~Di~Penta, F.~Fasano, R.~Oliveto, and A.~De~Lucia, ``A
  large-scale empirical study on the lifecycle of code smell co-occurrences,''
  {\em Information and Software Technology}, vol.~99, pp.~1--10, 2018.

\bibitem{nelson1976modified}
H.~E. Nelson, ``A modified card sorting test sensitive to frontal lobe
  defects,'' {\em Cortex}, vol.~12, no.~4, pp.~313--324, 1976.

\bibitem{guzzi2013communication}
A.~Guzzi, A.~Bacchelli, M.~Lanza, M.~Pinzger, and A.~v. Deursen,
  ``Communication in open source software development mailing lists,'' in {\em
  Proceedings of the 10th Working Conference on Mining Software Repositories},
  pp.~277--286, IEEE Press, 2013.

\bibitem{raudenbush2002hierarchical}
S.~W. Raudenbush and A.~S. Bryk, {\em Hierarchical linear models: Applications
  and data analysis methods}, vol.~1.
\newblock Sage, 2002.

\bibitem{de2018systematic}
E.~V. de~Paulo~Sobrinho, A.~De~Lucia, and M.~de~Almeida~Maia, ``A systematic
  literature review on bad smells---5 w's: which, when, what, who, where,''
  {\em IEEE Transactions on Software Engineering}, 2018.

\bibitem{azeem2019machine}
M.~I. Azeem, F.~Palomba, L.~Shi, and Q.~Wang, ``Machine learning techniques for
  code smell detection: A systematic literature review and meta-analysis,''
  {\em Information and Software Technology}, 2019.

\bibitem{chatzigeorgiou2010investigating}
A.~Chatzigeorgiou and A.~Manakos, ``Investigating the evolution of bad smells
  in object-oriented code,'' in {\em International Conference on the Quality of
  Information and Communications Technology}, pp.~106--115, IEEE, 2010.

\bibitem{spadini2018when}
D.~Spadini, M.~Aniche, M.-A. Storey, M.~Bruntink, and A.~Bacchelli, ``When
  testing meets code review: Why and how developers review tests,'' in {\em
  40th IEEE/ACM International Conference on Software Engineering},
  pp.~677--687, 2018.

\bibitem{conover}
W.~J. Conover, {\em Practical Nonparametric Statistics}.
\newblock Wiley, 3rd edition~ed., 1998.

\bibitem{cliff}
R.~J. Grissom and J.~J. Kim, {\em Effect sizes for research: A broad practical
  approach.}
\newblock Lawrence Erlbaum Associates Publishers, 2005.

\bibitem{ohtani2000bootstrapping}
K.~Ohtani, ``Bootstrapping r2 and adjusted r2 in regression analysis,'' {\em
  Economic Modelling}, vol.~17, no.~4, pp.~473--483, 2000.

\bibitem{nakagawa2013general}
S.~Nakagawa and H.~Schielzeth, ``A general and simple method for obtaining r2
  from generalized linear mixed-effects models,'' {\em Methods in Ecology and
  Evolution}, vol.~4, no.~2, pp.~133--142, 2013.

\bibitem{bland2000odds}
J.~M. Bland and D.~G. Altman, ``The odds ratio,'' {\em Bmj}, vol.~320,
  no.~7247, p.~1468, 2000.

\bibitem{murphy2012we}
E.~Murphy-Hill, C.~Parnin, and A.~P. Black, ``How we refactor, and how we know
  it,'' {\em IEEE Transactions on Software Engineering}, vol.~38, no.~1,
  pp.~5--18, 2012.

\bibitem{thongtanunam2015should}
P.~Thongtanunam, C.~Tantithamthavorn, R.~G. Kula, N.~Yoshida, H.~Iida, and
  K.-i. Matsumoto, ``Who should review my code? a file location-based
  code-reviewer recommendation approach for modern code review,'' in {\em
  Software Analysis, Evolution and Reengineering (SANER), 2015 IEEE 22nd
  International Conference on}, pp.~141--150, IEEE, 2015.

\bibitem{unkelos2015gamifying}
N.~Unkelos-Shpigel and I.~Hadar, ``Gamifying software engineering tasks based
  on cognitive principles: The case of code review,'' in {\em Cooperative and
  Human Aspects of Software Engineering (CHASE), 2015 IEEE/ACM 8th
  International Workshop on}, pp.~119--120, IEEE, 2015.

\bibitem{pantiuchina2018towards}
J.~Pantiuchina, G.~Bavota, M.~Tufano, and D.~Poshyvanyk, ``Towards just-in-time
  refactoring recommenders,'' in {\em Proceedings of the 26th Conference on
  Program Comprehension}, pp.~312--315, ACM, 2018.

\bibitem{sae2016context}
N.~Sae-Lim, S.~Hayashi, and M.~Saeki, ``Context-based code smells
  prioritization for prefactoring,'' in {\em Program Comprehension (ICPC), 2016
  IEEE 24th International Conference on}, pp.~1--10, IEEE, 2016.

\bibitem{vassallo2018context}
C.~Vassallo, S.~Panichella, F.~Palomba, S.~Proksch, A.~Zaidman, and H.~C. Gall,
  ``Context is king: The developer perspective on the usage of static analysis
  tools,'' in {\em 2018 IEEE 25th International Conference on Software
  Analysis, Evolution and Reengineering (SANER)}, pp.~38--49, IEEE, 2018.

\bibitem{baum2017optimal}
T.~Baum, K.~Schneider, and A.~Bacchelli, ``On the optimal order of reading
  source code changes for review,'' in {\em Software Maintenance and Evolution
  (ICSME), 2017 IEEE International Conference on}, pp.~329--340, IEEE, 2017.

\bibitem{spadini2019test}
D.~Spadini, F.~Palomba, T.~Baum, S.~Hanenberg, M.~Bruntink, and A.~Bacchelli,
  ``Test-driven code review: An empirical study,'' in {\em 2019 IEEE/ACM
  International Conference on Software Engineering (ICSE)}, pp.~1061--1072,
  IEEE, 2019.

\bibitem{pascarella2017classifying}
L.~Pascarella and A.~Bacchelli, ``Classifying code comments in java open-source
  software systems,'' in {\em Proceedings of the 14th International Conference
  on Mining Software Repositories}, pp.~227--237, IEEE Press, 2017.

\bibitem{pascarella2019classifying}
L.~Pascarella, M.~Bruntink, and A.~Bacchelli, ``Classifying code comments in
  java software systems,'' {\em Empirical Software Engineering}, pp.~1--39,
  2019.

\bibitem{sridhara2010towards}
G.~Sridhara, E.~Hill, D.~Muppaneni, L.~Pollock, and K.~Vijay-Shanker, ``Towards
  automatically generating summary comments for java methods,'' in {\em
  Proceedings of the IEEE/ACM international conference on Automated software
  engineering}, pp.~43--52, ACM, 2010.

\end{thebibliography}
